\documentclass[a4paper]{article}

\usepackage{amsmath,amsfonts,amssymb,amsthm}

\newtheorem{defi}{Definition}
\newtheorem{teo}{Theorem}

\newtheorem{prop}[teo]{Proposition}

\newtheorem{lem}[teo]{Lemma}
\newtheorem{rmk}{Remark }

\begin{document}
\title{The Tremblay-Turbiner-Winternitz system as extended Hamiltonian}
\author{Claudia Maria Chanu, 
 Luca Degiovanni, Giovanni Rastelli \\  Dipartimento di Matematica, \\ Universit\`a di Torino.  Torino, via Carlo Alberto 10, Italia.\\ \\ e-mail: claudiamaria.chanu@unito.it \\ luca.degiovanni@gmail.com \\ giovanni.rastelli@unito.it }

\maketitle

\begin{abstract}
We generalize the idea of ``extension of Hamiltonian systems'' -- developed in a series of previous articles --  which allows the explicit construction of Hamiltonian systems with additional non-trivial polynomial first integrals of arbitrarily high degree, as well as the determination of new superintegrable systems from old ones. The present generalization, that we call ``modified extension of Hamiltonian systems'', produces the third independent first integral for  the (complete) Tremblay-Turbiner-Winternitz (TTW) system, as well as for the caged anisotropic oscillator in dimension two.
\end{abstract}

\section{Introduction}
We further improve  a research started in \cite{CDR0} about a class of Hamiltonian systems admitting recursively computed, polynomial first integrals of high degree.
In \cite{CDRfi} we introduced a procedure that we called \emph{extensions of Hamiltonian systems},  lately generalized to \emph{$(m,n)$-extensions} in \cite{CDRraz}. Until now, our methods were not able to include  two of  the most important examples of two-dimensional superintegrable  Hamiltonians with high-degree first-integrals in whole generality (i.e., without setting some of the parameters appearing in the potential to be zero): the Tremblay-Turbiner-Winternitz (TTW) system \cite{TTW}
\begin{equation}\label{TTWo}
H=\frac 12 p_r^2+\frac{1}{r^2}\left(\frac 12p_\theta^2+\frac {\alpha_1}{\cos^2\lambda \theta}+\frac {\alpha_2}{\sin^2\lambda\theta}\right)+\omega r^2, \qquad \lambda \in \mathbb{Q},
\end{equation}
 and the two-dimensional caged anisotropic harmonic oscillator \cite{KKM}
 \begin{equation}
H=\frac 12 p_x^2+\frac 12 p_y^2-\omega^2(\lambda^2 x^2+y^2)+\frac b{x^2}+\frac c{y^2},\qquad \lambda \in \mathbb{Q}. 
 \end{equation}
  In the present paper, we slightly modify the procedure for building the first integral in the $(m,n)$-extensions to include both the above-mentioned  systems. The "modified $(m,n)$-extensions" that we introduce here still keep the relevant features of the extension method, namely: the high-degree first integrals are explicitly determined, the extension procedure can be applied to $n$-degrees of freedom systems and it is linked to their geometry, the modified procedure remains a powerful tool for the creation of new superintegrable systems from old ones, as we plan to show in a future paper that will generalise \cite{CDRsuext}.
  
\section{$(m,n)$-extensions and the TTW Hamiltonian}
We adopt the notation introduced in \cite{CDRraz} (dropping the tildes for simplicity). Let $L$ be a  Hamiltonian
on a $d$-dimensional Poisson manifold $Q$. For any pair of positive integers $m$, $n$,  we denote by $H_{m,n}$ its $(m,n)$-extension,  that is the function
\begin{equation}\label{mn_ext}
H_{m,n} = \frac{1}{2}p_u^2+\frac{m^2}{n^2}\alpha(u) L+\frac{m^2}{n^2} \beta(u), 
\end{equation}
defined on the $(d+2)$-dimensional Poisson manifold $T \times Q$ where $T$ has canonical symplectic form $\mathrm{d}p_u\wedge \mathrm{d}u$, and the functions $\alpha$ and $\beta$ are listed in
Table 1 below.
In \cite{CDRgen} (for $n=1$) and in \cite{CDRraz} (for any $n\in \mathbb{N}-\{0\}$) we proved that,
if there exist constants $c$ and $L_0$ (not both zero) such that the equation
\begin{equation}\label{e1}
X_L^2(G)=-2(cL+L_0)G,
\end{equation}
admits a solution $G$ on $Q$ ($X_L$ denotes the Hamiltonian vector field of $L$), then a first integral of $H_{m,n}$ is given by
\begin{equation}\label{mn_int}
K_{m,n}=U_{m,n}^m(G_n)=\left(p_u+\frac{m}{n^2} \gamma(u)   X_L\right)^m(G_n)
\end{equation}
where $G_n$ is the $n$-th term of the recursion
\begin{equation}\label{rec}
G_1=G, \qquad G_{n+1}=X_L(G)\,G_n+\frac{1}{n}G\,X_L(G_n), 
\end{equation}
 and $\gamma(u)$ is given in Table 1.

\begin{table}[h]
\begin{center}
\begin{tabular}{|c|c|c|}
\hline
&$ c=0$ & $ c \neq 0$ \cr

\hline

$ \alpha = -{\gamma}'= $&$A$ &  $\dfrac { c \vphantom{\frac 12} }{S_\kappa^2( c u)}$ \cr 

$\beta = L_0 {\gamma}^2=$ &$\vphantom{\dfrac 12} L_0A^2u^2$ & $0$  \cr

$ \gamma = $& $-Au$ & $\dfrac 1{T_\kappa( c u)}$\cr  

\hline
\end{tabular}
\end{center}
\caption{Functions involved in the $(m,n)$-extension of $L$}
\end{table}

We remark that 
\begin{itemize}
\item
if (\ref{e1}) has a solution $G$ for $c\neq 0$, then  we may assume without loss of generality
$L_0=0$;
\item
in Table 1, $A$ and $\kappa$ are arbitrary constants and
the functions $S_\kappa$ and $T_\kappa$ are the trigonometric tagged functions
$$
S_\kappa(x)=\left\{\begin{array}{ll}
\frac{\sin\sqrt{\kappa}x}{\sqrt{\kappa}} & \kappa>0 \\
x & \kappa=0 \\
\frac{\sinh\sqrt{|\kappa|}x}{\sqrt{|\kappa|}} & \kappa<0
\end{array}\right.
\qquad
C_\kappa(x)=\left\{\begin{array}{ll}
\cos\sqrt{\kappa}x & \kappa>0 \\
1 & \kappa=0 \\
\cosh\sqrt{|\kappa|}x & \kappa<0
\end{array}\right.,
$$
$$
T_\kappa(x)=\frac {S_\kappa(x)}{C_\kappa(x)},
$$
(see \cite{CDRraz} for a summary of their properties).

\item
in \cite{CDRgen} (where a slightly different notation was adopted) it is shown that the functions (\ref{mn_ext}) and (\ref{mn_int})   Poisson-commute only for the $\alpha$, $\beta$, $\gamma$  appearing in Table 1.
\end{itemize}

The method of exensions can be effectively applied in two ways:  it allows to construct new superintegrable systems from known superintegrable ones with one less degree of freedom (see \cite{CDRsuext}), or
it can be a constructive proof  that a given Hamiltonian  $H_\lambda$, depending on a  parameter $\lambda=m/n$, admits a first integral of degree depending on $\lambda$, provided we are able to write $H_\lambda$ as the $(m,n)$ extension of a Hamiltonian $L$ admitting a solution for (\ref{e1}). In \cite{CDRgen} we give a geometric characterization  of the natural Hamiltonians $H$ that can be written  as the extension of a Hamiltonian $L$ with one less degree of freedom.

In both types of applications the key point is equation (\ref{e1}).
For a natural Hamiltonian $L$ on the cotangent bundle of a Riemannian $N$-dimensional 
manifold $M$, this equation has been studied in the case of $G$ independent of the momenta
(see \cite{CDRPol,CDRfi}) and linear in the momenta (\cite{CDRgen}).
For natural Hamiltonians, the recursion (\ref{rec}) provides solutions $G_n$ of (\ref{e1})  polynomial in the momenta with degree depending on $n$, starting from any known solution $G$, possibly independent from the momenta.
 
When $N=1$ and  $G$ is a polynomial in $p$ of  degree $r$ i.e.,
$$
L=\frac 12 p^2+V(q),\quad G=\sum\limits_{i=0}^{r}\eta_i(q)p^i,
$$
the equation  (\ref{e1}) is equivalent to
\begin{equation}\label{e2}
\left\{ \begin{array}{ll}
\eta''_i+c\eta_i=0, & i=r-1,r,\\
\, \\
\eta''_{i-2}+c\eta_{i-2}=(2i+1)V'\eta_i'-(2cV+2L_0-iV'')\eta_i, & i=r-1,r,\\
\, \\
\eta''_{i-2}+c\eta_{i-2}= (2i+1)V'\eta_i'-(2cV+2L_0-iV'')\eta_i-\\
-(i+1)(i+2)(V')^2\eta_{i+2},& 1<i\leq r-2,\\
\, \\
(2i+1)V'\eta_i'-(2cV+2L_0-iV'')\eta_i=
\\ (i+1)(i+2)(V')^2\eta_{i+2}, &
  i=0,1,\\
\end{array} \right.
\end{equation}
where $\eta_j=0$ for $j<0$ or $r<j$. We remark that the $\eta_i$ with indices even and odd are not appearing together into (\ref{e2}). Since we are interested in a particular solution $G$,  it is not restrictive to assume that the  solutions $G$ of (\ref{e2}) are polynomials containing    even (resp.\ odd) powers of $p$ only when $r$ is even (resp.\ odd).
Moreover, when $V$ is given, the system can be solved iteratively, starting from $\eta_r$ to $\eta_0$ (resp. $\eta_1$). The (\ref{e2}.4) gives a further condition on $\eta_0$ (resp. $\eta_1$) for $r$ even (resp. odd) that makes the system generally  unsolvable.

For $G$ linear and homogeneous in $p$, the equations (\ref{e2}) become
\begin{equation}\label{gl}
\left\{ \begin{array}{l}
\eta''+c\eta=0, \\
3V'\eta'+ \eta V''-2\eta (cV+L_0)=0.
\end{array} \right.
\end{equation}

The solution of the first equation is
\begin{equation}\label{eta}
\eta_r=a_1S_{c}(q)+a_2C_{c}(q).
\end{equation}
then, the second equation can be solved for $V(q)$  to determine all the possible potentials in $L$ admitting an extension for $G=\eta p$. The possible solutions are summarized in Table 2.
\begin{table}[h]
\begin{center}
\begin{tabular}{|c|c|c|}
\hline
&$V$ & $ G=\eta p$ \cr

\hline
&&\cr

$c\neq 0 $ &  $\dfrac{c_1 + c_2 \eta'}{\eta^2}-\dfrac{L_0}c$ &  $\Big(a_1S_{c}(q)+a_2C_{c}(q)\Big)p $\cr 

$c=0$, $a_1\neq 0$ &$\dfrac{L_0}{4 a_1}\eta^2+\dfrac{c_1}{\eta^2}+c_2$ & $(a_1q+a_2)p$  \cr

$ c=0$, $a_1=0$ & $L_0q^2+c_1q+c_2$ & $a_2p$\cr  

\hline
\end{tabular}
\end{center}
\caption{Solutions of system (\ref{gl}).}
\end{table}

The solution of the first equation is (\ref{eta}), then, the second equation can be solved for $V(q)$  to determine all the possible potentials in $L$ admitting an extension for $G=\eta p$. The solution of the last equation is, since (\ref{gl}) holds,
\begin{equation}\label{VV}
V=\frac{c_1 + c_2 \eta'}{\eta^2}+\frac{L_0}2\frac{F(q)}{\eta'},
\end{equation}
where $F'=\eta$ and $c(F''+cF)=0$ for $\eta'$ not identically zero. If $\eta'=0$, then  we have necessarily $c=0$ and
$$
V=L_0q^2+c_1q+c_2.
$$
In the case $c=1$, we have $\beta=0$ and the $(m,n)$ extension (\ref{mn_ext}) of $L$ is
$$
H=\frac 12 p_u^2+\frac{m^2}{n^2S_\kappa^2(u)}\left(\frac 12p^2+V\right).
$$
Moreover, up to ineffective translations in $q$, we get $G=(\sin q)p$ and  (\ref{VV}) becomes, up to additive constants,
\begin{equation}\label{Vv}
V=\frac{c_1+c_2\cos q}{\sin^2q}.
\end{equation}

\begin{prop}\label{ttw2}
The TTW Hamiltonian without harmonic term
\begin{equation}\label{ttwe}
H=\frac 12 p_r^2+\frac 1{r^2}\left(\frac 12 p_\theta^2+\frac{\alpha_1}{\cos^2\lambda \theta}+ \frac{\alpha_2}{\sin^2\lambda\theta}\right),\quad \lambda=\frac mn,
\end{equation}
is the $(2m,n)$-extension, with $c=1$, $L_0=0$ and $\kappa=0$,  of
\begin{equation}\label{LTTW}
L=\frac 12 p_q^2+\frac{c_1+c_2\cos q}{\sin^2q},
\end{equation}
with 
\begin{equation} \label{par}
q=2\lambda \theta, \qquad c_1=\frac {\alpha_1+\alpha_2}{2\lambda^2}, \qquad c_2=\frac {\alpha_2-\alpha_1}{2\lambda^2}.
\end{equation}
\end{prop}

\proof
Since the potential of (\ref{LTTW}) is of the form (\ref{Vv}), 
for $c=1$ and $L_0=0$,
the Hamiltonian $L$ admits the $(2m,n)$-extension 
$$
H=\frac 12 p_u^2+\frac{4m^2}{n^2S_\kappa^2(u)}\left(\frac 12 p_q^2+\frac{c_1+c_2\cos q}{\sin^2q}\right),
$$
which, for $\kappa=0$, becomes
$$
H=\frac 12 p_u^2+\frac{4m^2}{n^2u^2}\left(\frac 12 p_q^2+\frac{c_1+c_2\cos q}{\sin^2q}\right).
$$
Through the coordinate transformation $q=2\lambda \theta$, $u=r$ and with $c_1$ and $c_2$ given by (\ref{par}), we get (\ref{ttwe}).
Vice versa, to any $(m,n)$-extension of (\ref{LTTW}) corresponds (\ref{ttwe}) with $\lambda =\frac{m}{2n}$.
\qed

\section{Modified extensions}
For any $m,n,s,k\in \mathbb N-\{0\}$, 
let us consider the functions $\alpha$, $\beta$, $\gamma$, the recursion $G_n$, and the operator  $$U_{m,n}=p_u+\frac m{n^2}\gamma X_L$$ introduced in (\ref{e1}), (\ref{rec}), and in Table 1 for  the $(m,n)$-extension  $H_{m,n}$ of $L$. 
We define the functions
\begin{eqnarray}
\bar H_{m,n}=\frac 12 p_u^2+\frac {m^2}{n^2}\alpha L+ \frac {m^2}{n^2}\beta +\omega \gamma^{-2}, \qquad \omega\in \mathbb{R}.
\label{ee1}
\end{eqnarray}
\begin{equation}
\bar K_{2s,k}=\left(U_{2s,k}^2+2\omega \gamma^{-2}\right)^sG_k.\label{ee2}
\end{equation}
The main result of this paper is

\begin{teo} \label{teo1} 
For  $m=2s$, 
  we have
$$
\{\bar H_{2s,n},\bar K_{2s,n}\}=0.
$$
For $m=2s+1$,
$$
\{\bar H_{2s+1,n},\bar K_{4s+2,2n}\}=0.
$$
\end{teo}
The proof of Theorem \ref{teo1} depends on the following Lemma.

\begin{lem}\label{lem_conti}
Let us consider the Hamiltonian 
\begin{equation}
H=\frac 12 p_u^2+f(u)+\left(\frac {2m}n\right)^2\alpha(u) L, \qquad m,n\in \mathbb{N}-\{0\},
\end{equation}
where the Hamiltonian $L$ does not depend on $(u,p_u)$, and the operator $W$ defined by
\begin{equation}
W(G)=\left(p_u+\frac{2m}{n^2}\gamma(u) X_L\right)^2(G)+(2f(u)+h(u))G.
\end{equation}
We have $\{H,W^m(G)\}=0$ iff
\begin{eqnarray}
 & & X_L^2(G)=-2n^2(cL+L_0)G,\label{eqG}\\
  & & \gamma''+2c\gamma'\gamma=0, \label{eqg}\\
 & & \alpha(u)=-\gamma', \label{eqa}\\
 & & f(u)=  \frac{4m^2}{n^2}L_0\gamma^2+
 \frac{f_0}{\gamma^2}+\frac{h_0}{2},
 \label{eqf}\\
 & & h(u)= -\frac{8m^2}{n^2}L_0\gamma^2-h_0, \label{eqh}
\end{eqnarray}
for some constants $c$, $L_0$ not both zero.
\end{lem}

\begin{proof} We follow the same path of the proof of Proposition 1 of \cite{CDRfi}. 
It is well known that, if two operators $A$, $B$ satisfy
$$
[[A,B],B]=0,
$$
where $[,]$ denote operator commutation, then
$$
AB^m(G)=B^{m-1}\left(m[A,B]+BA\right)(G),
$$
for any function $G$. If $B$ is injective, then  $AB^m=0$ iff $m[A,B]+BA=0$. When $A$ and $B$ are the Hamiltonian vector field $X_{\bar H}$ of $\bar H_{2m,n}$ and $U_{2m,n}^2+2\omega \gamma^{-2}$ respectively,  the equation $AB^m=0$ is equivalent to the statement of the Theorem.
First, we  check that $[[X_H,W],W]=0$:   we have
\begin{eqnarray*}
[X_H,W]=4\frac m{n^2}\gamma'p_u^2+\left(h'-8\frac{m^2}{n^2}\alpha'L\right)p_u-4\left(\frac m{n^2}f'\gamma+4\frac {m^3}{n^4}\alpha' \gamma L\right)X_L+\\
+8\frac{m^2}{n^4}p_u\gamma \gamma'X_L^2,
\end{eqnarray*}
and therefore the condition $[[X_H,W],W]=0$ is evidently satisfied. 
For $G(q^i,p_i)$ such that  $X_L(G)\neq 0$ $W$ is injective and
\begin{eqnarray*}
&& m[X_H,W]G+WX_HG=p_u^24\frac {m^2}{n^2}(\gamma'+\alpha)X_LG+\\
&& \qquad +p_u\left( \left(mh'-8\frac{m^3}{n^2}\alpha'L\right)G+8\frac {m^3}{n^4}\gamma\left(\gamma'+2\alpha\right)X_L^2G\right)+\\
&&\qquad +4\frac {m^2}{n^2}\left(\alpha(h+2f)-f'\gamma-4\frac {m^2}{n^2}\alpha'\gamma L\right)X_LG+16\frac {m^4}{n^6}\alpha\gamma^2X_L^3G,
\end{eqnarray*}
which is a polynomial in $p_u$. By requiring that  the coefficient of $p_u^2$ vanishes, under the non-restrictive assumption $\gamma'\neq 0$, we obtain (\ref{eqa}).
The requirement that the coefficient of $p_u$ vanishes, by separating terms in $u$ from those in $q^i,p_i$, is equivalent to (\ref{eqG}), (\ref{eqg}) and
\begin{eqnarray*}
h'+16\frac{m^2}{n^2}L_0\gamma\gamma'=0,
\end{eqnarray*}
where $c$, $L_0$ are separation constants, whose integration gives (up to an additive constant) (\ref{eqh}).
By imposing (\ref{eqa}),  the coefficient of degree $0$ in $p_u$ becomes
\begin{equation}\label{conti}
16\frac {m^4}{n^6}\gamma'\gamma^2X_LG\left(-\frac {n^4}{4\gamma'\gamma^2m^2}\left(\gamma'(h+2f)+f'\gamma\right)+\frac {n^2\gamma''}{\gamma \gamma'}L-\frac {X_L^3G}{X_LG}\right).
\end{equation}
By (\ref{eqG}) and (\ref{eqg}), we have 
$$\frac {X_L^3G}{X_LG}=-2n^2(cL+L_0), \qquad \frac {\gamma''}{\gamma \gamma'}=-2c,
\qquad 
\frac {n^2}{4\gamma^2m^2}(h+h_0)=-2L_0
$$ and the bracket (\ref{conti}) reduces to
$$
-\frac {n^4}{4\gamma'\gamma^2m^2}\left(2\gamma'f- \gamma' h_0+f'\gamma\right)+4n^2L_0,
$$
which vanishes if and only if
$$
f'+2\frac{\gamma'}{\gamma}f-h_0\frac{\gamma'}{\gamma}- 16L_0\frac{m^2}{n^2}\gamma\gamma'=0 
$$
i.e., iff (\ref{eqf}) holds.
We remark that  $2f+h=f_0\gamma^{-2}$. Thus, we have   $W(G)=U_{2m,n}^2(G)$ only for $f_0=0$ i.e., when  $f=(2m/n)^2 \beta$, up to an additive constant.
\end{proof}

\begin{proof} of Theorem \ref{teo1}.
We apply Lemma \ref{lem_conti} to $H=\bar H_{2s,n}$ and $W=U_{2s,n}^2+2\omega \gamma^{-2}$.
Then, we get  that the functions are in involution iff $\alpha,$ $\beta$ and $\gamma$ are the functions  appearing in Table 1. Moreover, we have, up to inessential additive constants, $f(u)=  \frac {4s^2}{n^2}\beta +\omega \gamma^{-2}$ and $2f(u)+h(u)=\omega \gamma^{-2}$. 
Since $H_{m,n}=H_{2m,2n}$ the second part follows directly.
\end{proof}

Hence, we can give the following definition:

\begin{defi} \rm
We call the Hamiltonian $\bar H_{m,n}$ admitting  the first integral  $\bar K_{2s,n}$, if $m=2s$,or $\bar K_{2m,2n}$,
if $m=2s+1$, the \emph{modified extended Hamiltonian} of $L$. 
\end{defi}

For $\omega=0$, we drop back to $(m,n)$-extensions.

We determine now an explicit expression for the first integrals.

\begin{lem} For $r\leq m$, $\Lambda=-2(cL+L_0)$, we have

\begin{equation}\label{EEcal}
U_{m,n}^r(G_n)= P_{m,n,r}G_n+D_{m,n,r}X_{L}(G_n),
\end{equation}
with
$$
P_{m,n,r}=\sum_{k=0}^{[r/2]}\binom{r}{2k}\, \left(\frac mn  \gamma \right)^{2k}p_u^{r-2k}\Lambda^k,
$$
$$
D_{m,n,r}=\frac 1{n}\sum_{k=0}^{[(r-1)/2]}\binom{r}{2k+1}\, \left(\frac mn  \gamma \right)^{2k+1}p_u^{r-2k-1}\Lambda^k, \quad m>1,
$$
where $[\cdot]$ denotes the integer part and $D_{1,n,1}=\frac 1{n^2} \gamma$.
\end{lem}
\begin{proof}
It follows directly from the analogous expressions in \cite{CDRraz}.
\end{proof}

\begin{prop} The expansion of the first integral (\ref{ee2}) is
\begin{equation}
\bar K_{2m,n}=\sum_{j=0} ^{m}\binom{m}{j}\left(\frac {2\omega}{\gamma^2}\right)^jU_{2m,n}^{2(m-j)}(G_n).
\end{equation}
\end{prop}
\begin{proof} Since  $U_{2s,n}^2$ and $2\omega \gamma^{-2}$ commute as operators, the $s$-th power in (\ref{ee2}) coincides with  the power of a binomial.
\end{proof}

\begin{teo}
Let $\{L_1=L, \ldots, L_k\}$ be a set of functionally independent first integrals of the  Hamiltonian $L$ on $Q$. If $\bar H_{2s,n}$, $\bar K_{2s,n}$ determine a non-trivial modified $(m,n)$-extension of $L$, then $\{\bar H_{2s,n}, \bar K_{2s,n}, L_1, \ldots, L_k\}$ are all functionally independent.
\end{teo}
\begin{proof}
i) The rank of the Jacobian matrix of the $(\bar H_{2s,n}, \bar K_{2s,n},L_1,\ldots,L_k)$ w.r. to the coordinates $(u,p_u,q^i)$, where $i=1,\ldots, k$ and $q^i$ denote here both momenta and configuration coordinates, is equal to the rank of the square $(k+2)\times(k+2)$ matrix
\begin{eqnarray}
J=
\left(\begin{matrix} \frac {\partial \bar H_{2s,n}}{\partial u}& \frac {\partial \bar H_{2s,n}}{\partial p_u} &\frac {4m^2}{n^2}\alpha \frac{\partial L}{\partial q^1} & \vdots & \frac {4m^2}{n^2}\alpha \frac{\partial L}{\partial q^k} \cr
\frac {\partial \bar K_{2s,n}}{\partial u} &\frac {\partial \bar K_{2s,n}}{\partial p_u} & \frac {\partial \bar K_{2s,n}}{\partial q^1} & \vdots & \frac {\partial \bar K_{2s,n}}{\partial q^k} \cr 0 & 0 & \frac {\partial L}{\partial q^1} & \vdots & \frac {\partial L}{\partial q^k}  \cr 0 & 0 & \frac {\partial L_2}{\partial q^1} & \vdots & \frac {\partial L_2}{\partial q^k}  \cr
\vdots & \vdots & \vdots &\vdots & \vdots\cr
0 & 0 & \frac {\partial L_{k}}{\partial q^1}  & \vdots & \frac {\partial L_{k}}{\partial q^k}
\end{matrix}\right)
\end{eqnarray}
 where the  order of the $q^i$  is chosen so that the rank of the $k\times k$ submatrix in the bottom-right corner, which we denote by $J_k$, is $k$. The determinant of $J$ is given by
$$
\det(J)=\left(\frac {\partial \bar H_{2s,n}}{\partial u}\frac {\partial \bar K_{2s,n}}{\partial p_u}-\frac {\partial \bar H_{2s,n}}{\partial p_u}\frac {\partial \bar K_{2s,n}}{\partial u}\right)\det (J_k).
$$
Therefore, because $\det (J_k)\neq 0$ by assumption,
$\det (J)=0$ iff $$\frac {\partial \bar H_{2s,n}}{\partial u}\frac {\partial \bar K_{2s,n}}{\partial p_u}-\frac {\partial \bar H_{2s,n}}{\partial p_u}\frac {\partial \bar K_{2s,n}}{\partial u}=0.$$ Since $\{\bar H_{2s,n}, \bar K_{2s,n}\}=0$ and $\alpha \neq 0$, it follows  $\{L, \bar K_{2s,n}\}_{(q^i)}=0$. In the last equation the highest-degree term in $p_u$ is $p_u^{2s}\{L,G_n\}=p_u^{2s}X_L(G_n)$ and as we assumed $X_L(G_n)\neq 0$.
\end{proof}

\section{Examples}
\subsection{The TTW family}\label{TTW}

The complete TTW system in the Euclidean plane with $\lambda=\frac mn$ is
$$
H=\frac 12 p_u^2+\frac{m^2}{n^2u^2}\left(\frac 12p_\psi^2+\frac {\alpha_1}{\cos^2\psi}+\frac {\alpha_2}{\sin^2\psi}\right)+\omega u^2.
$$
From Proposition \ref{ttw2} we have that $H$  coincides with
$$
\bar H_{2m,n}=\frac 12 p_u^2+\frac{4m^2}{n^2u^2}\left(\frac 12p_q^2+\frac{c_1+c_2\cos q}{\sin^2q}\right)+\omega u^2,
$$
where $q=2\psi$ and $c_1=\frac {\alpha_1+\alpha_2}{2\lambda^2}$ $c_2=\frac {\alpha_2-\alpha_1}{2\lambda^2}$. For $\lambda=1$ the corresponding first integral of $\bar H_{2,1}$ is
\begin{eqnarray*}
\bar K_{2,1}=p_qp_u^2 \sin q+\frac 4up_up_q^2 \cos q-\frac 4{u^2}p_q^3\sin q+\frac 4u p_u\frac {c_2(\cos ^2q+1)+2c_1\cos q}{\sin ^2 q}+\\
+\frac 2{u^2}p_q\frac {\omega u^4\sin ^2q-4(c_1+c2 \cos q)}{\sin q}.
\end{eqnarray*}

\subsection{The  two-dimensional anisotropic caged oscillator}\label{cage}
If we put $c=0$, $V=aq^2+b/q^2$ into (\ref{gl}) we get $\eta= q$ and $a=L_0/4$. The modified $(m,n)$-extension of $L=\frac 12 p_q^2+V$ is therefore
$$
\bar H_{m,n}=\frac 12 p_u^2+\frac {m^2}{n^2}A\left(\frac 12 p_q^2+\frac {L_0}4q^2+\frac b{q^2}\right)+\frac{m^2}{n^2}L_0A^2u^2+\frac \omega{A^2u^2}.
$$
By putting $A=1$, $x=\frac nm q$, we have
$$
\bar H_{m,n}=\frac 12 p_u^2+\frac 12 p_x^2+L_0\frac {m^2}{n^2}(\frac {m^2}{4n^2}x^2+u^2)+\frac b{x^2}+\frac \omega{u^2},
$$
that is the generic two-dimensional superintegrable anisotropic caged oscillator \cite{KKM}.

\begin{rmk}\rm
It is easy to see that $\bar H_{1,1}$ is the same for  examples \ref{TTW} and \ref{cage} (it is enough to pass from Cartesian to polar coordinates), but the rational parameter appears in two different ways, leading to different types of extensions.
\end{rmk}

\section{Conclusions}
With the inclusion of the TTW system and the caged anisotropic oscillator into the scheme of extensions, we somehow complete a work started in \cite{CDR0} about systems admitting polynomial first integrals of high degree. Through several papers, our ``extension'' approach showed several  unexpected possibilities, such as the creation of new superintegrable systems from old ones \cite{CDRsuext},  the connection with warped manifolds theory \cite{CDRraz}. With the present article, we gave an essay of the flexibility of our approach, whose possible developments, also for quantum systems, will be analysed in future.

\end{document}